\documentstyle[draft,psfig]{mn}

\def\etal{{\rm et al.\ }}

\def\gs{\mathrel{\raise0.35ex\hbox{$\scriptstyle >$}\kern-0.6em 
\lower0.40ex\hbox{{$\scriptstyle \sim$}}}}
\def\ls{\mathrel{\raise0.35ex\hbox{$\scriptstyle <$}\kern-0.6em 
\lower0.40ex\hbox{{$\scriptstyle \sim$}}}}

\def\rcool{r_{\rm cool}}
\def\fcool{f_{\rm cool}}
\def\tcool{t_{\rm cool}}
\def\esn{\epsilon_{\hbox{sn}}}

\def\MNRAS{MNRAS}
\def\ApJ{ApJ}

\def\keV{\,\hbox{keV}}
\def\ergs{\,\hbox{ergs}\,\hbox{s}^{-1}}

\def\erg{\,\hbox{erg}}

\title[Galaxy Formation and the X-ray Evolution of Clusters]
    {The Impact of Galaxy Formation on the X-ray Evolution of Clusters}
\author[Bower, Benson, Baugh, Cole, Frenk \& Lacey]
{
R.G. Bower$^1$, A.J. Benson$^1$, C.G. Lacey$^2$, C.M. Baugh$^1$, S. Cole$^1$,
\& C.S. Frenk$^1$\\
 $^1$ Department of Physics, University of Durham, South Road, 
		Durham DH1 3LE, UK\\
 $^2$ SISSA, via Beirut, 2-3, 34014 Trieste, Italy\\
}

\begin{document}

\label{firstpage}

\maketitle

\begin{abstract}
We present a new model for the X-ray properties of the intracluster medium
that explicitly includes heating of the gas by the energy released during
the evolution of cluster galaxies. We calculate the evolution of clusters
by combining the semi-analytic model of galaxy formation of Cole et
al. with a simple model for the radial profile of the intracluster gas. We
focus on the cluster X-ray luminosity function and on the relation between
X-ray temperature and luminosity (the T-L relation). Observations of these
properties are known to disagree with predictions based on scaling
relations which neglect gas cooling and heating processes. We show that
cooling alone is not enough to account for the flatness of the observed T-L
relation or for the lack of strong redshift evolution in the observed X-ray
luminosity function. Gas heating, on the other hand, can solve these two
problems: in the $\Lambda$CDM cosmology, our model reproduces fairly well
the T-L relation and the X-ray luminosity function. Furthermore, it
predicts only weak evolution in these two properties out to $z=0.5$, in
agreement with recent observational data. A successful model requires an
energy input of 1--2 $\times 10^{49}$ ergs per solar mass of stars
formed. This is comparable to the total energy released by the supernovae
associated with the formation of the cluster galaxies. Thus, unless the
transfer of supernovae energy to the intracluster gas is very (perhaps
unrealistically) efficient, additional sources of energy, such as
mechanical energy from AGN winds are required. However, the amplification
of an initial energy input by the response of the intracluster medium to
protocluster mergers might ease the energy requirements. Our model makes
definite predictions for the X-ray properties of groups and clusters at
high redshift. Some of these, such as the T-L relation at $z\simeq 1$, may
soon be tested with data from the Chandra and Newton satellites.
\end{abstract}

\begin{keywords}
galaxies: formation 
\end{keywords}

\section{Introduction}
One of the fundamental puzzles of the X-ray universe concerns the
relation between the X-ray luminosity and gas temperature of
clusters of galaxies. A simple scaling analysis (Kaiser 1986) suggests that the
temperature and luminosity should be related by $T\propto L^{1/2}$.
Temperatures have now been measured for the diffuse X-ray emission for
an extensive range of groups and clusters (David et al. 1993; Ponman
et al. 1996; Allen \& Fabian 1998; Markevitch 1998; Mulchaey \&
Zabludoff 1998; Arnaud \& Evrard 1999; Helsdon \& Ponman 2000). In
contrast to the theoretical prediction, the observations
show a much shallower trend, approximately $T\propto L^{1/3}$.

A closely related problem is the evolution of the cluster X-ray
luminosity function.  Kaiser's (1986) analysis of the evolution of the
X-ray properties of clusters suggested that dense, X-ray luminous
associations of galaxies should be more numerous in the intermediate and high
redshift universe.  This possibility was soon ruled out by the initial
results of the EMSS cluster survey (Gioia et al. 1990; Henry et
al. 1992), which quickly established that clusters in the distant
universe have a comparable space density to those of the local universe. This
has been confirmed in more recent ROSAT surveys (eg., Jones et al. 2000).

Initially, one might have hoped that including radiative cooling of
the gas (omitted from Kaiser's analysis) might resolve this
discrepancy. Unfortunately, it is extremely difficult to obtain
numerically convergent results from simulations once gas cooling is
included.  The
difficulty is inherent to the problem. Because the universe is dense
at early times, cooling is initially very efficient. This leads to an
unrealistically large fraction of the halo baryon content cooling at
high redshifts to
form very small galaxies. As White \& Rees
(1978), White \& Frenk (1991), Cole (1991), Suginohara \& Ostriker
(1998) and Pearce et al.\ (2000) amongst others have shown, some form
of heating is required to overcome this catastrophe. In Appendix~A, we
examine how the cooled gas fraction depends on cluster mass, and find
that the cooled fraction depends too weakly on cluster temperature to
explain the discrepancy. Therefore radiative cooling cannot by itself
solve the problem with the temperature-luminosity relation (see Bryan et al.\ 
2000 and Balogh et al.\ 2001 for an extended discussion).

One approach to this problem that has given encouraging results is to
assume that the gas is ``preheated'' before collapsing into the
cluster (Evrard \& Henry 1991; Kaiser 1991; Navarro, Frenk \& White
1995). This creates an entropy floor in the gas, ensuring that it
remains at low densities in low mass systems, and results in a much
improved match to the T-L relation (Balogh, Babul \& Patton 1999;
Valageas \& Silk 1999; Tozzi \& Norman 2000). This model also provides
an encouraging match to the surface brightness profiles of low mass
groups (Ponman et al. 1999). The problem with this model is in
explaining the origin of this diffuse heating and its apparent
uniformity.

Prolonged heat input from galaxy formation has been suggested as a
solution by Wu et al. (1998, 1999a) and Cavaliere et al. (2000). They
calculated the response of the gas profile to the energy input from
supernovae by using a simple energetic approach. They start by assuming a
1-parameter form for the gas profile or equation of state and then
calculate how the total energy of the halo gas depends on this 
parameter. By then solving for the dependence of the parameter on 
the energy excess relative to an initial profile
corresponding to the case of no heat input, the contributions to the
energy balance from gravity, radiative cooling and supernova heating
can be taken into account in calculating the new gas
distribution. This approach successfully accounts for the shallow
present-day temperature-luminosity (T-L) relation if galaxy formation
has a roughly uniform efficiency in haloes of different masses. Since
the binding energy per particle increases with halo mass, while the
additional heating remains roughly constant, high mass clusters are
almost unaffected while the gas in low mass groups becomes unbound.

In this paper, we develop a new model for the evolution of the masses
and density profiles of the hot gas in galaxy, group and cluster
haloes, using the semi-analytic galaxy formation scheme of Cole et
al. (2000) to predict the evolution of the supernova
heating rate in dark haloes of different masses, and calculating the
response of the gas profile to this heating using a method related to
those of Wu et al. and Cavaliere et al. The semi-analytic scheme is an
elaboration of that described by Baugh et al.\ (1998), and is based on
similar principles to the models described by Kauffmann, White \&
Guiderdoni (1993) and reviewed by Somerville \& Primack (1999).  We
apply our model to study the evolution of the X-ray luminosity
function and the temperature-luminosity relation.

The structure of this paper is as follows.  Our method for relating
the gas distribution in the halo to the supernova energy input is
presented in \S2. The predicted X-ray properties are detailed in
\S3. In section \S3.1, we show that supernova heating is able to
produce the observed slope and normalization of the present-day T-L
relation only if the efficiency with which the supernova explosion
energy is transferred to the diffuse intracluster medium (ICM) is very
high, or if the stellar initial mass function (IMF) is tilted to produce an
overabundance of high-mass stars relative to the IMF in the solar
neighbourhood. Alternatively, heat input from AGN activity may be
required. In \S3.2, we apply this model to the X-ray luminosity
function of galaxy clusters. We compare the evolution predicted by the
model within a flat, $\Omega_0=0.3$, CDM cosmology with the available
observations of intermediate redshift clusters. In \S3.3, we consider
the X-ray properties of the universe at very high redshifts, and in
\S3.4, we compare the expectations based on our galaxy formation model
with those from two extreme models for the redshift evolution of the
heat input.  A further discussion of the problems and a restatement of
our conclusions are given in \S4 and \S5.

\section{The Model}

Wu et al. (1998, 1999a) have suggested a simple approach that allows
non-gravitational heating to be incorporated into the calculation of
the properties of cluster gas. Starting from a default distribution,
gas is redistributed to larger and larger radii until the total
energy increase matches the energy input from galaxy formation.  The
effect of heat input may affect the distribution of gas within
clusters in a variety of ways.  Our approach differs from that of Wu et al.\
both in the way we determine the default gas distribution and in the way
we modify the gas distribution in response to the excess energy
input. Firstly, while Wu et al.\ adopt a complex prescription for the default 
gas distribution based on the cluster's 
gravitational binding energy, our default profile is based explicitly
on the observed properties of high temperature rich clusters. We are
able to do this because the ranges of excess energy that we consider
have little impact on the gas distribution in these systems. 
Secondly, Wu et al.\ explore a variety of models in which heating occurs
either by uniformly varying the gas temperature, or by varying the
polytropic index of the gas. In contrast, our approach is empirical and
motivated by the observations of Arnaud \& Evrard (1999) and Lloyd-Davies,
Ponman \& Cannon (2000) who find that the gas profiles of clusters become
systematically shallower at lower temperatures. We therefore assume that
the overriding effect of heating is to reduce the slope of the radial
density profile of the gas. Our empirical approach does not require us to
choose explicitly between the isothermal and polytropic regimes. Instead,
for our given density profile, we solve for hydrostatic equilibrium in
order to determine the gas temperature.

We assume that the dark matter density of the halo follows a Navarro,
Frenk \& White (1997, NFW) profile:
\begin{equation}
\rho_{\rm DM}(r) = \frac{\rho_s a^3}{r(r+a)^2}
\end{equation}
The density normalization $\rho_s$ and dependence of scale radius $a$
on halo mass are calculated as described in Cole et
al. (2000). We parameterize the gas distribution using a conventional
$\beta$ model (Cavaliere \& Fusco-Femiano 1976):
\begin{equation}
\rho_{\rm gas}(r) = \frac{\rho_c}{(1+(r/r_c)^2)^{3\beta/2}}
\end{equation}
Both of these distributions are assumed to apply within the virial
radius $R_{\rm vir}$, where $R_{\rm vir}$ is calculated from a
spherical collapse model as described by Cole et al.  The first step
is to fix the parameters of the default radial gas profile that applies in
the absence of any energy input from supernovae.  We initially
distribute the gas with a core radius $r_c$ that is a fixed fraction (7\%)
of the virial radius $R_{\rm vir}$, and set $\beta=0.7$ in order to match
observations of the most massive clusters (eg. Lloyd-Davies, Ponman \&
Cannon 2000).  The temperature of the gas at the virial radius is set
to $0.6 T_{\rm vir}$, as suggested by the numerical simulations of Eke,
Navarro \& Frenk~(1998) and Frenk et al.~(1999). Here the virial
temperature is defined as
\begin{equation}
T_{\rm vir} = \frac{1}{2} \frac{\mu m_H}{k} \frac{G M}{R_{\rm vir}}
\label{eq:tvir}
\end{equation}
where $M$ is the total mass within $R_{\rm vir}$, $\mu$ is the mean
molecular mass (we take $\mu=0.59$ for fully ionized gas) and $m_H$ is
the mass of a hydrogen atom. The temperature of the gas at smaller
radii is then found by solving for hydrostatic equilibrium in the
gravitational potential of the dark matter.  This technique accurately
reproduces the luminosity weighted temperature of the cluster
simulated by Frenk et al. (1999).  We adjust the normalization
$\rho_c$ of the default gas profile so that the baryonic mass fraction
(ie.  gas plus galaxies) enclosed within the virial radius is equal to
the cosmic baryon fraction. Treating the total mass in this way takes
into account the effect of cooled gas and stars that are locked into
galaxies, thus reducing the hot gas fraction of the cluster. X-ray
luminosities are calculated from the gas within the cluster virial
radius, since material at larger radii is unlikely to be in
hydrostatic equilibrium. In practice, this cut-off has little
influence on the X-ray luminosity, since that is dominated by the
densest material in the cluster core.

\begin{figure}
\vbox{
\begin{centering}
 \psfig{file=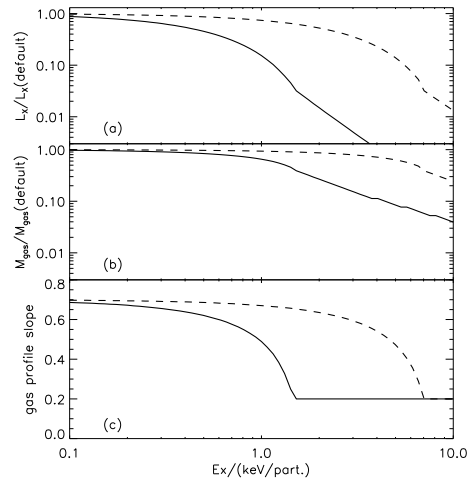,width=8.5cm}
\end{centering}
}
\caption{Panel (a): the dependence of the X-ray luminosity of a
cluster on the excess energy injected into the ICM. $E_{\rm x}$ is the
injected energy per baryon of the hot gas. The luminosity is plotted
relative to the luminosity of the default profile. The two clusters
shown have virial temperatures of 1 keV (solid line) and 5 keV (dashed
line). The kink at $L_{\rm X}/L_{\rm X}(\hbox{default})\sim0.03$
corresponds to the minimum allowed $\beta$-slope of 0.2.  Larger
excess energies are accommodated by increasing the temperature of the
gas. The fraction of the default gas mass remaining within the cluster
virial radius is shown in Panel (b); while Panel (c) shows the dependence
of the slope of the gas density profile, $\beta$, on the excess energy.}
\label{fig:exlxfig}
\end{figure}

Having established the default profile for a given cluster, we reduce
the slope $\beta$ of the gas profile while keeping the core radius
$r_c$ constant, until the total energy (thermal plus gravitational) of
the gas is increased by the required amount.  As the profile changes,
we keep the pressure and density (and thus temperature) at the cluster
virial radius fixed at the value found for the default profile. This
results in the mass of gas within the virial radius being less than
for the default profile. The excess gas is assumed to be expelled. It is
displaced to the virial radius and included in the energy balance
calculation, but not in the calculation of the X-ray luminosity. The
temperature of this material is assumed to be the same as that of the
gas at $R_{\rm vir}$. This corresponds to the lowest plausible
temperature for the expelled gas to be both in pressure equilibrium
with its surroundings and buoyant with respect to the remaining
cluster material.  

We have chosen the virial radius as the point at which to normalize
our density profiles because this approximately delineates the region
of the cluster that is in dynamical equilibrium and separates it from
the outer parts of the cluster that are dominated by bulk
inflow. Outside the virial radius, the gas is unlikely to be in
hydrostatic equilibrium. Close to the virial radius, the infalling gas is
shocked so that its bulk motion is converted to internal energy. In
the one dimensional 
simulations of Knight \& Ponman (1997), where the infalling material
has uniform initial entropy, the shock radius occurs at 1-$1.5
R_{\rm vir}$, in line with the boundary radius we assume here.
In
three dimensional simulations, the shock radius is more poorly defined
because the infalling material already has a range of initial
entropies and this tends to smooth out the shock, but the same general
picture applies. The gas pressure at the virial radius is thus
regulated by the dynamical pressure of the infalling gas.

There is a limit to the overall energy increase that can be
accommodated by flattening the gas profile. Rather than letting $\beta$
become arbitrarily low, we impose a minimum value $\beta_{\rm
min}=0.2$. If the required slope falls below this value, we set $\beta =
\beta_{\rm min}$, and instead allow the temperature of the gas (both
inside and outside $R_{\rm vir}$) to rise to accommodate the excess
energy, giving up the condition $T(R_{\rm vir})=0.5 T_{\rm vir}$, but
maintaining the condition that $T=T(R_{\rm vir})$ for the expelled
gas. Since the pressure at $R_{\rm vir}$ is kept fixed, the gas
density must then fall and a greater fraction of the gas mass is
expelled. Our results are not sensitive to the exact choice of
$\beta_{\rm min}$, since the total energy of the cluster depends only
very weakly on $\beta$ for $\beta<0.4$. The lowest values in observed
systems are $\beta\sim 0.35$ (Lloyd-Davies, Ponman \& Cannon 2000).

Figure~\ref{fig:exlxfig} shows the relation between energy input (ie.
the excess energy) per baryon, $E_{\rm x}$, and X-ray luminosity
$L_{\rm X}$ for clusters with virial temperatures of 1 and 5~keV
(Panel (a)), and the fraction of the original gas mass that remains
within the cluster virial radius (Panel (b)). Note that the decline in
X-ray luminosity is much more rapid than the decline in the gas mass
within $R_{\rm vir}$.  Experimenting with different schemes for
modelling the effects of heating, such as keeping the mass within
$R_{\rm vir}$ constant, shows that the fixed pressure assumption is
the most effective at reducing the X-ray luminosity for a given energy
input. For comparison, Panel (c) shows the dependence of the slope 
parameter $\beta$ on the injected energy.

The effects on the temperature-luminosity relation of radiative
cooling are discussed in Appendix~A. We argue there that radiative
cooling will cause some flattening of the T-L relation
compared to the case of no cooling and no energy input, but not enough
by itself to match the observed relation. This is because, although
the fraction of gas inside the cooling radius increases with
decreasing halo mass, this dependence is too weak in the relevant mass
range to flatten the relation to $T \propto L^{1/3}$.

X-ray luminosities and luminosity-weighted temperatures for individual
dark matter haloes are calculated using Peacock's (1996) analytic fit
to the Raymond-Smith cooling function, assuming a metal abundance of
$1/3$ solar. The results change by only 
a few percent if we use the tabulated cooling function directly. 
This includes both
bremsstrahlung and recombination processes and is adequate for the
range of halo masses considered here. Representative halo formation
and merging histories are generated using a Monte-Carlo method based
on the extended Press-Schechter model as described by Cole et
al.~(2000).  This ensures that our model includes the correct halo
mass distribution and assigns collapse redshifts to individual
haloes. We use the properties of the halo at its collapse time for
determining the X-ray properties. 

We adopt the cosmological parameters $\Omega_0=0.3$, $\Lambda_0=0.7$,
$h=0.7$, $\sigma_8=0.8$ and  $\Gamma=0.19$, where $\Lambda_0$ is the
cosmological constant measured in units of $3 H^2_0/c^2$, $\sigma_8$
is the linear theory mass variance in spheres of radius $8h^{-1}$Mpc
at the present, and $\Gamma$ is the shape parameter of the initial
spectrum of density fluctuations defined by
Efstathiou, Bond \& White (1992). With these parameters, our model
X-ray temperature function matches the data of Eke et al. (1996, 1998). 
Note our value of $\sigma_8$ differs slightly from that inferred by Eke et al.\
because our 
luminosity-weighted gas temperatures are $\sim$15\% higher than the
cluster virial temperatures that they use.  This temperature offset is
consistent with the results of hydrodynamical simulations of clusters
(eg. Frenk et al.  2000), and depends on the profile adopted for the
gas distribution in the central regions of the clusters (which
dominate the X-ray luminosity).  In order to match the observed
temperature function, we have lowered $\sigma_8$ from 0.93 to 0.80. We
retain the $\Gamma=0.19$ power spectrum shape preferred by Eke et al.

We normalize the model to fit the observed temperatures and
luminosities of the most luminous X-ray clusters by adjusting the
cosmic baryon fraction. These clusters are almost unaffected by
energy injection. Their predicted luminosities scale as $L_X^{\rm model}
\propto \Omega_b^2 h (1+z_f)^{3/2} T^2$ for given values of $\Omega_0$ and
$\Lambda_0$ (see Appendix~A), while the observationally inferred
values scale as $L_X^{\rm obs} \propto h^{-2}$. Matching models to
observations, we find that $\Omega_{\rm b}=0.025 h^{-3/2}$ gives a
good fit to the observed X-ray luminosities of clusters with virial
temperatures greater than $7\keV$. Once the model is normalized in
this way, the gas fractions within $1.5 h^{-1}$~Mpc are consistent
with more detailed observational estimates.

We calculate the amount of energy injected into the halo gas using the
detailed semi-analytic model of galaxy formation of Cole et al. (2000,
{\sc GALFORM}). This model follows the formation of dark matter halos
through hierarchical clustering and merging, using merger histories
generated by a Monte Carlo scheme, and then calculates the cooling and
collapse of gas within halos to form galaxies, and the formation of
stars from the cool gas. The model includes the effects of
feedback from supernova explosions in expelling cold gas from galaxy disks, 
and traces the mergers of galaxies within
common dark matter halos. From the star formation history of each
galaxy, we can then calculate the rate of supernovae as a function of
time, once we assume an IMF. The model therefore predicts for each
dark halo both the total mass of baryons in galaxies, and the total
number of supernovae that have occurred in galaxies in that halo and in
its progenitor halos at earlier times. We calculate the excess energy
of the hot gas in a particular halo by summing the energies of all the
supernovae that have occurred in the progenitors of the halo up to its
formation time, after allowing for some efficiency for this energy to
be transferred to the halo gas. Both the excess energy and the
fraction of baryons in galaxies have a systematic variation with halo
mass and a random scatter in haloes of a given mass due to different
cooling and star formation histories.

We ran the GALFORM model with the same parameter values (for star
formation, feedback, etc) as in Cole et
al. (2000). These values were
chosen to match observed properties of present-day galaxies, in
particular luminosity functions, colours and stellar mass-to-light
ratios. The GALFORM model does not include the effects of the
modification of the halo gas profile due to energy injection when it
calculates the rate of gas cooling, so our modelling is not fully
self-consistent. A fully self-consistent treatment would require us to
reconsider the form of the star formation law and investigate afresh
what combination of parameters gives the best fit of predicted to
observed galaxy properties in the present-day universe, once we
include the effects of energy injection on gas cooling. This
self-consistent treatment is postponed to a future paper. In the
present paper, we have a more limited aim, which is to investigate the
consequences for the X-ray properties of the ICM of including energy
injection based on a specific ab initio model of galaxy formation
which has already been shown to reproduce a wide range of
observational data on galaxy properties. In a self-consistent
calculation, the effect of injection will be to reduce the amount of
gas which cools onto galaxies, mimicking the effect of reducing
$\Omega_b$. For this reason, we allow the value of $\Omega_b$ used in
GALFORM to be smaller than the value used in calculating the ICM
properties. Specifically, GALFORM was run with $\Omega_b=0.02$, as in
Cole et al. (2000), to calculate galaxy masses and supernova
rates. The total baryonic mass was calculated assuming $\Omega_{\rm
b}=0.025 h^{-3/2}$ ($\Omega_b=0.043$ for $h=0.7$), and the ICM mass
calculated as the difference of the total baryon mass and the mass in
galaxies.

We treat the energy injected into the ICM per unit mass of stars formed as
a free parameter, which we will adjust in order to fit the present-day form
of the T-L relation. We adopt the parameterisation that an energy $\esn
10^{49} \erg$ goes into heating the ICM per M$_\odot$ of stars
formed. There are two sources of uncertainty in trying to predict the value
of $\esn$ from first principles: (a) The number of Type~II supernovae per
$M_{\odot}$ of stars formed, $\eta_{sn}$, depends on the IMF and on the
minimum stellar mass $m_{sn}$ for core-collapse. For a Salpeter IMF with an
upper mass limit of 125M$_\odot$, lower mass limit of 0.1 M$_\odot$ and
$m_{sn}= 8M_{\odot}$, $\eta_{sn}=0.007$. The Cole et al. (2000)
semi-analytic model uses a somewhat different IMF, that of Kennicutt
(1983), which is a better fit to that observed in the solar neighbourhood
at $m< M_{\odot}$, but this predicts the same number of supernovae per unit
mass as in the Salpeter case, once the fraction of brown dwarfs (with
$m<0.1 M_{\odot}$) is normalized to match the observed galaxy luminosity
function as in Cole et al.  A higher SN rate applies if the IMF is skewed
towards high mass stars, or if the lower mass limit for the progenitors of
supernovae is reduced (e.g. Chiosi et al.  1992). Lower supernova rates are
suggested by recent analyses of the metal abundance of the intracluster
medium (Renzini, 1997, Kravtsov \& Yepes, 2000). (b) Each Type~II supernova
explosion releases an energy of about $10^{51}$ ergs (eg. Woosley \& Weaver
1986), but some fraction of this energy is lost by radiative cooling as the
remnant is expanding into the interstellar medium of the host galaxy, and
so is not available to heat the ICM. For instance, Thornton et al.\ (1998)
find that 70-90\% of the energy is lost in this way. Thus, for a Salpeter
IMF, we predict $\esn=0.7$ if none of the supernova energy is lost by
radiation, but in practice a much smaller value seems likely.

We can convert the heating efficiency $\esn$ into an excess energy per
baryon $E_{\rm X}$ once the fraction of baryons converted into stars
$f_{\rm gal}$ is known. For clusters with $1 < T < 10 {\rm keV}$,
our models give $f_{\rm gal}\approx 0.13$ for $h=0.7$, with only a
weak dependence on $T$ in this range. This value scales with the
Hubble constant roughly as $f_{\rm gal}\approx 0.16 h^{1/2}$, if the
parameters in the semi-analytic model are adjusted to match observed
galaxy luminosities and mass-to-light ratios at each $h$, resulting in
the stellar mass scaling as $h^{-1}$, and if the total baryon fraction
is scaled as $\Omega_b \propto h^{-3/2}$ to match the X-ray
luminosities.  The excess energy per baryon in the ICM is then
\begin{equation}
E_{\rm X} \approx 0.50\esn 
\left(f_{\rm gal}\over 0.16 h^{1/2}\right) h^{1/2} \hbox{keV per particle} 
\end{equation}
The value of $\esn$ that is required to make the model clusters fit
the observed T-L relation then scales with $h$ approximately as
$h^{-1/2}$, since a certain ($h$-independent) value of $E_{\rm X}$ is
required at each $T$ to shift the clusters onto the observed relation
from the relation that applies in the absence of heating or cooling.

As explained above, we expect the heating efficiency $\esn$ due to
supernovae to be significantly less than unity, if the IMF has the
standard form and radiative losses are significant.  However,
additional energy may be available from active galactic nuclei. AGNs
may emit mechanical energy in the form of winds or jets that can heat
the gas in the 
surrounding dark halo. Although the details of the fuelling of AGN
activity are unclear (see Nulsen \& Fabian 2000, for a recent
discussion), the requirements for this fuelling are similar to those
for star formation, and the two processes may be closely linked.  We
will assume that the AGN activity parallels the star formation
activity in the galaxies. If all galaxies harbour black holes with
masses close to those suggested by Magorrian et al.\ (1998), we can
estimate the available energy as follows. If we assume that the 
total energy released by the formation of each black hole of mass 
$M_{\rm BH}$ is approximately $0.1 M_{\rm BH}c^2$ (eg., Rees 1984). 
Magorrian's relation suggests $M_{\rm BH} \sim 0.06 M_{\rm
stars}$ where $M_{\rm stars}$ is the mass in stars (strictly, the
bulge mass). Combining these relations shows that the available energy
is $\sim 10^{52}$ ergs per $M_\odot$ of stars, or $\esn=1000$,
compared to $\esn\sim 1$ from supernovae.  Thus, the energy
contribution to the ICM from AGN could, in principle, exceed that from
galaxies by several orders of magnitude (depending on the fraction of
the AGN luminosity emitted as kinetic energy). For this reason we will
allow for the possibility that $\esn>1$.

\section{Results}

\subsection{The Temperature-Luminosity Relation}

\begin{figure*}
\begin{centering}
\psfig{file=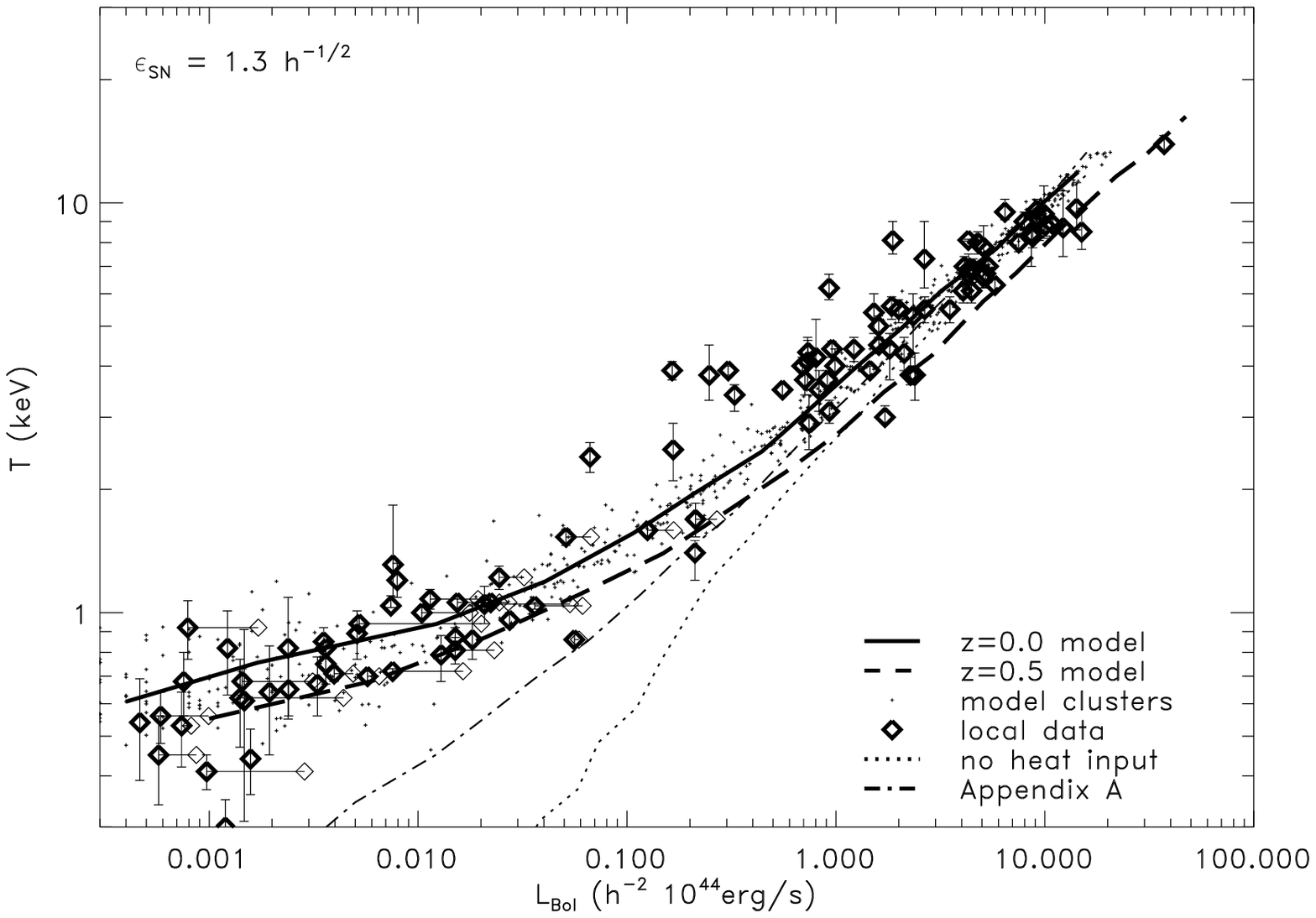,width=14cm}
\psfig{file=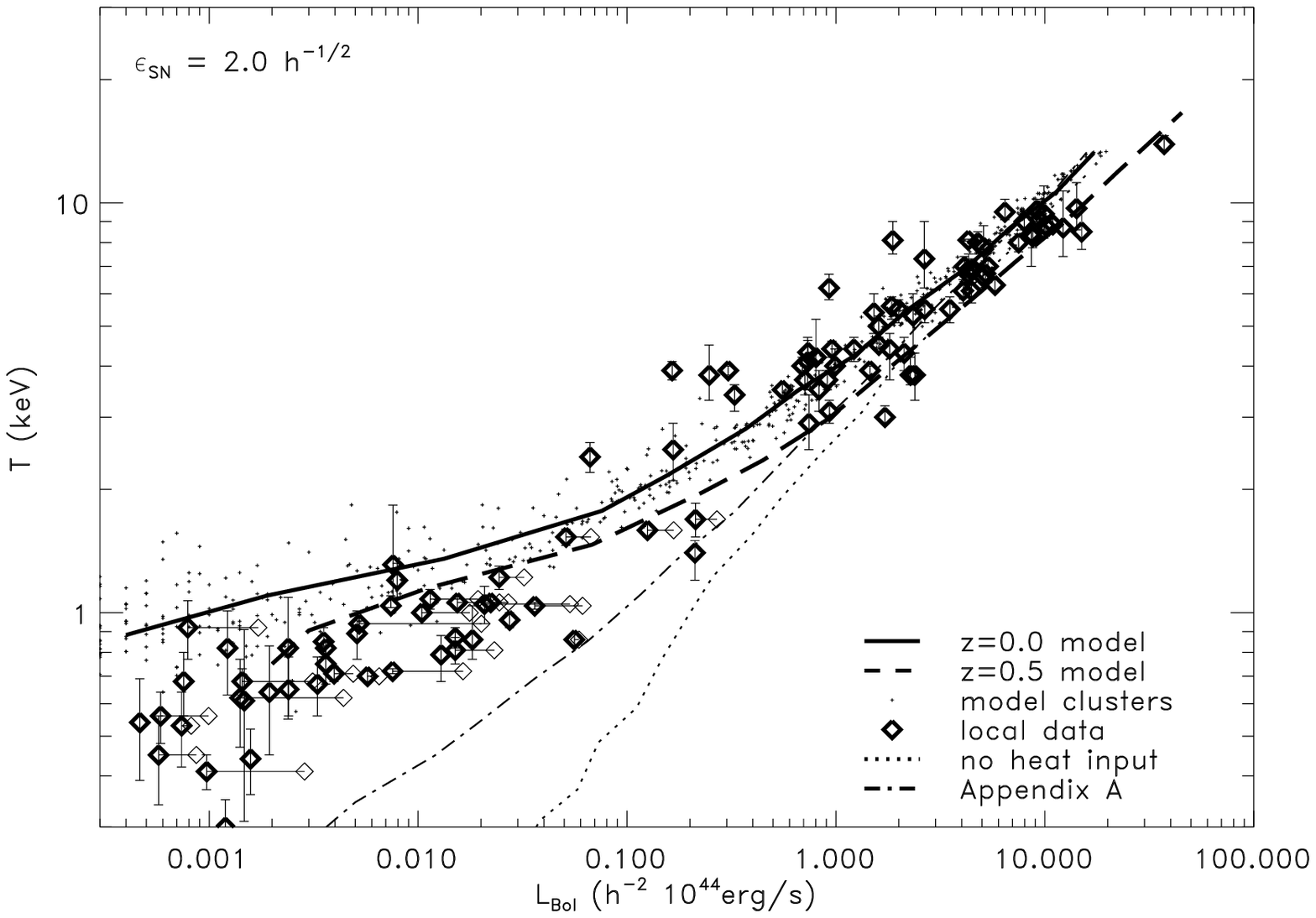,width=14cm}
\end{centering}
\caption{A comparison of the predicted and observed
T-L relations for  heating efficiencies of $\esn=1.3\, h^{-1/2}$ {\it
(upper panel)} and $\esn=2\, h^{-1/2}$ {\it
(lower panel)}. The 
distribution of model clusters at $z=0$ is shown as points,
with the thick solid line showing the median T at each L. The thick
dashed line 
shows the median T-L relation in this model at $z=0.5$. Bold diamonds
are observational data points for clusters and groups 
with $z<0.1$ taken from a variety of sources as described in the
text; lighter diamonds illustrate the effect of the aperture
correction recommended by Helsdon et al. (2000). 
 The dotted line shows the median T-L relation from a
model with $\esn=0$, i.e. in which heat input from galaxy formation is
turned off.  The dot-dashed line shows an estimate of the T-L relation when 
there is no heat input from galaxies, but 
additional gas is removed following the proceedure described in Appendix~A.}   
\label{fig:ltfig}
\end{figure*}

As expected, if the ICM is assumed to be heated only by gravitational
collapse, with no energy injection from galaxies, then the model
clusters fail to match the observed slope of the T-L relation. Data
from David et al. (1993) show a slope close to $T \propto L^{1/3}$, a
result that has been confirmed by the analysis of more recent ASCA
observations (Arnaud \& Evrard 1999). Although, the brightest clusters may
follow a steeper slope than this when the luminosities are corrected
for contributions from cooling flows (Markevitch 1998; Allen \& Fabian
1998), the $L^{1/3}$ slope extends down to groups of much lower
luminosity (Ponman et al. 1996; Mulchaey \& Zabludoff 1998; Helsdon \&
Ponman 2000).  The X-ray properties of our model clusters are compared
with observational data in Figure~\ref{fig:ltfig}, in which the dotted 
line shows the predicted median relation for the case when there is no
energy injection. We prefer to plot this relation with temperature on
the vertical axis as the observational uncertainties are far greater
for X-ray temperatures than luminosities.

In order to match the observed form of the T-L relation, it is
necessary to introduce very substantial heating of the ICM. In the
upper panel of Fig~\ref{fig:ltfig}, we show the T-L relation at $z=0$
in a model with a heating efficiency $\esn=1.3 h^{-1/2}$. (Note that
we calculate all our models for $h=0.7$, but then assume the scaling
of $\esn$ with $h$ that was derived in the previous section.) This 
value of $\esn$ is already larger than can be accounted for by
supernova feedback alone, if the IMF has the conventional solar
neighbourhood form, even if there are no radiative energy losses. This
suggests that a significant contribution from AGN is probably also
required.  If the heating produced by galaxies is smaller than this,
the model predictions at the bright end fall too steeply with
decreasing luminosity. Even with an efficiency of $\esn=1.3 h^{-1/2}$,
the predicted T-L relation seems somewhat too steep for the most
luminous clusters ($L_{\rm X}>10^{44} h^{-2}\ergs$). These clusters
are not much affected by this amount of heating and tend to follow the
self-similar slope. To bring the most luminous clusters into line with
the observed T-L slope requires that the energy injection efficiency
be increased to $\esn=2.0 h^{-1/2}$. However, the model then fails to
reproduce the presence of X-ray luminous clusters with temperatures
below 1~keV (Fig.~\ref{fig:ltfig}, lower panel). Thus, it seems that
the heating efficiency needs to be slightly greater in the progenitor
haloes of the most massive clusters, in order to produce the best
match to the T-L relation.  This might be the case if galaxy formation
and/or AGN activity were even more strongly biased to high density
regions than in the Cole et al.\ model.

The model results show considerable scatter around the median
relation, which arises from two 
sources. Firstly, haloes collapse over a range of redshifts, leading to
some variation in core density. Secondly, the efficiency of galaxy
formation varies from halo to halo, resulting in considerable scatter
in the level of heating. The scatter in the model is in reasonably
good agreement with the observational data, although it fails to
encompass a small number of clusters with high temperature and low
X-ray luminosity. The transient effects of cluster mergers are not
included in the present model.

The free parameters of the model have now been fixed to match the
present-day T-L relation, and so the evolution to higher redshift provides
a test of the model. As discussed in the previous section, the
evolution of the T-L relation is determined by a competition between
the increasing density of collapsed structures, the temperature
distribution of the clusters and the relative importance of the excess
energy.  The thick dashed line in Fig.~\ref{fig:ltfig} shows the
predicted median T-L relation at $z=0.5$. There is little evolution in
this relation, consistent with presently available data on distant
clusters (Mushotzky \& Scharf 1997; Fairley et al. 2000). There is a
tendency in the model for clusters of a given temperature to become
more X-ray luminous at high redshift, but the trend is too weak to be
rejected on the basis of currently available data. Fairley et
al. (2000) have analysed the evolution of the T-L relation in a large
sample of clusters from $z=0.2$ to 0.8.  They fit their results to a
parameterised form, $L \propto T^{3.15} (1+z)^{\eta}$, and derive
$\eta = 0.60\pm0.38$ for an {\it open} $\Omega_0=0.3$ universe. This
corresponds to $\eta=0.75\pm 0.48$ in our flat cosmology, since the
luminosities inferred from the data are then greater. At $T=5\keV$ our
model produces a factor of 1.86 increase in the median cluster
luminosity over the redshift interval 0.0 to 0.5, corresponding to
$\eta=1.54$. Thus, the evolution predicted by our model is compatible
(at $1.6\sigma$) with that observed by Fairley et al.

\subsection{The X-ray Luminosity Function}

\begin{figure}
\begin{centering}
  \psfig{file=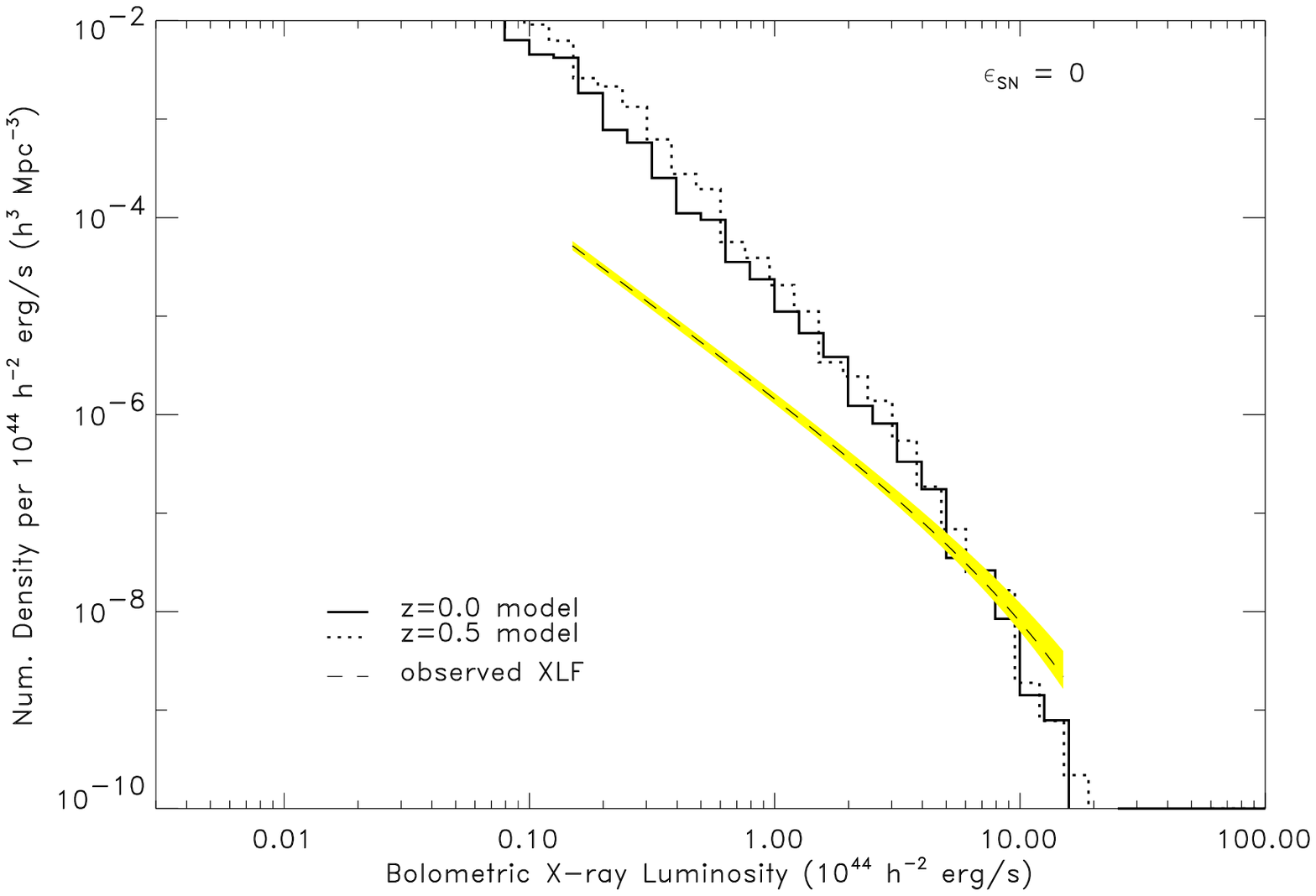,width=8cm}
  \psfig{file=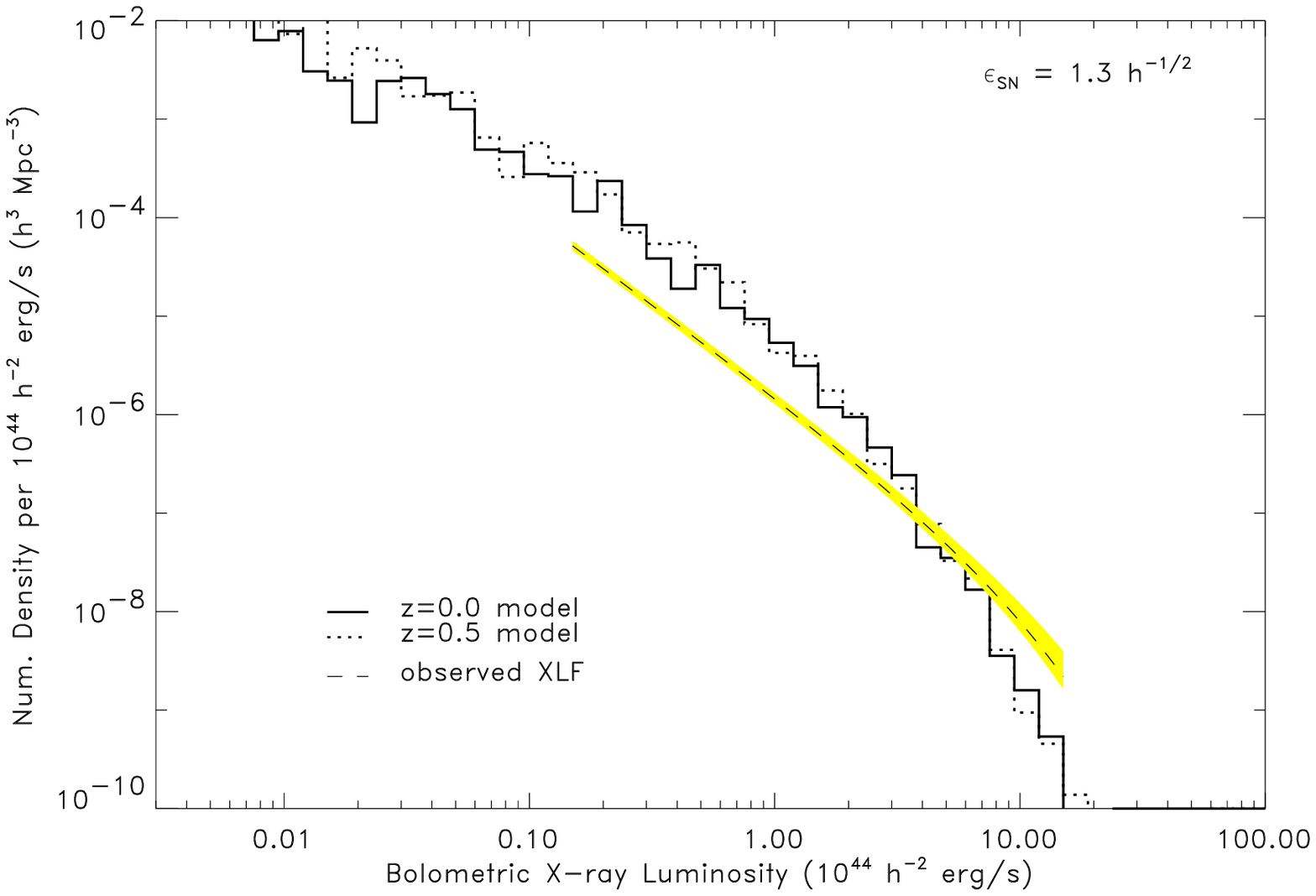,width=8cm}
  \psfig{file=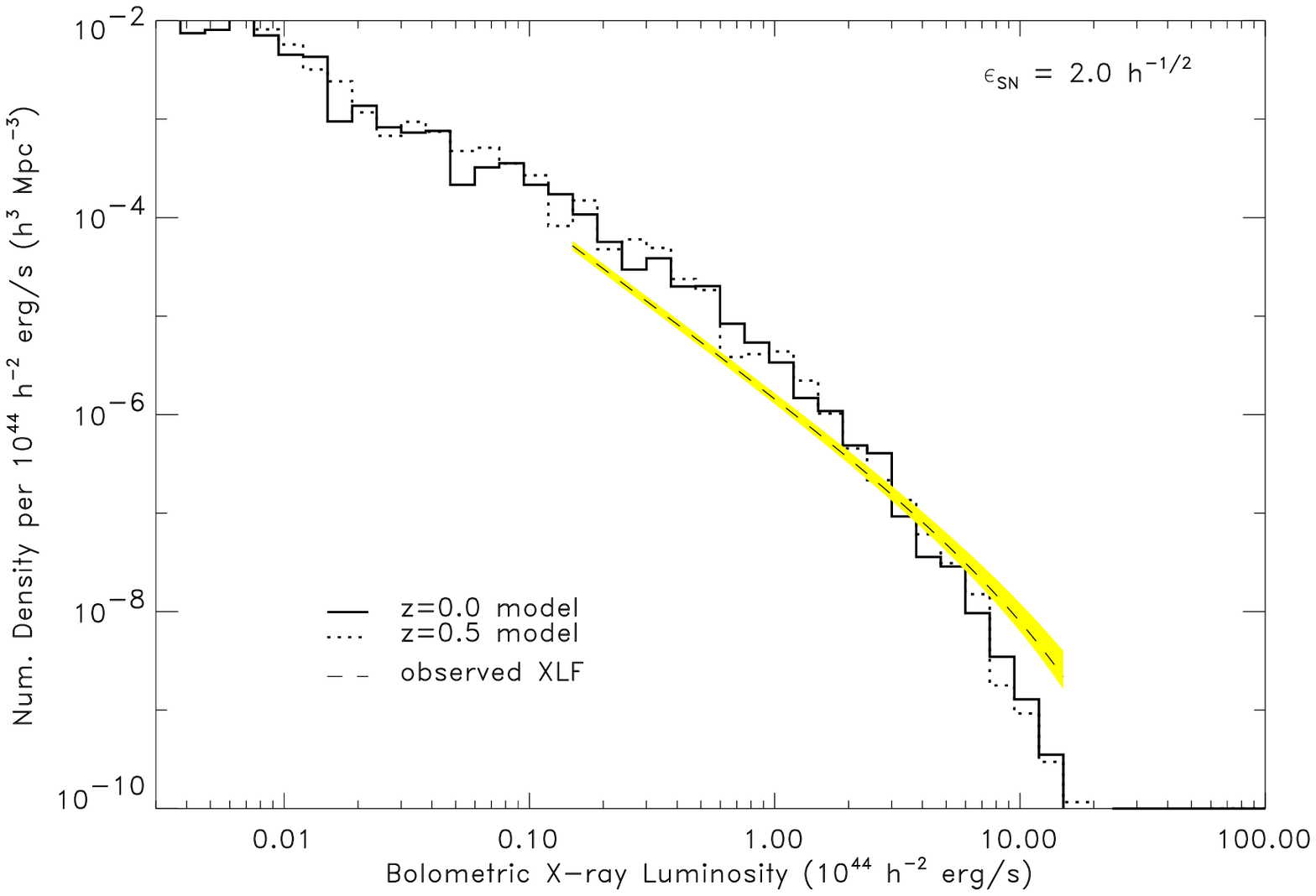,width=8cm}
\end{centering}
\caption{{\it Upper panel:} the X-ray luminosity function (XLF) at
$z=0$ (solid) and at $z=0.5$ (dotted) for the case when there is no heating
($\esn=0$).  The dashed line shows the observed present-day
luminosity function of Ebeling et al.\ (1997), with the shaded region
illustrating the statistical uncertainty. {\it Middle Panel:} the luminosity
function derived for the case $\esn=1.3 h^{-1/2}$; lines as in the previous 
panel. {\it Lower panel:} as above, but for $\esn=2.0 h^{-1/2}$.}
\label{fig:xlffig}
\end{figure}

The heating model provides a good description of the present-day T-L
relation, and can account for its observed lack of evolution. We now
consider the X-ray luminosity function (XLF). Since the galaxy
formation model generates a statistical sample of haloes, the X-ray
luminosity function can be readily calculated by combining the
different halos with appropriate weights.  In Fig.~\ref{fig:xlffig} we
show the predicted luminosity functions at $z=0$ (solid line) and
$z=0.5$ (dotted line). We show the 
luminosity function that is derived without heating (ie., $\esn=0$)
in the top panel. The middle and lower panels correspond to the
values of the efficiency parameter, $\esn=1.3 \,h^{-1/2}$ and
$\esn=2\,h^{-1/2}$ respectively, chosen to match the observed
temperature-luminosity relation.  These predictions are compared to the
observed local luminosity function derived by Ebeling et al (1997)
from the ROSAT all-sky ``Bright Cluster'' survey (BCS). 

Without heating, the model is an extremely poor fit to the observed data:
this is expected since we have chosen the CDM power spectrum to match
the observed cluster temperature function. In this case the luminosities
of low temperature haloes are too high, and this is reflected in the 
luminosity function which is too steep.
Since the available XLF data are restricted to relatively bright clusters, we
expect to obtain the best fit with $\esn=2.0\,h^{-1/2}$ rather than with
$\esn=1.3\,h^{-1/2}$. This is indeed the case, although even for
$\esn=2.0\,h^{-1/2}$ the model luminosity function is still somewhat too
steep. The discrepancy can be traced back to the slight bend in the
T-L relation seen in Fig.~\ref{fig:ltfig}, at the temperature at which
the effect of the injected energy becomes significant. The fit could
be fine-tuned by making the energy input increase more
strongly with halo mass (eg. if galaxy formation were more efficient
in proto-cluster regions), or by adjusting the cosmological
parameters. For example, adopting $\sigma_8=0.73$ and $\Gamma=0.07$
reduces the number of small mass haloes while boosting the abundance
of the highest mass objects. This gives a significantly improved match
to the luminosity function, but such a small value of $\Gamma$ may not
be compatible with measurements of large-scale galaxy clustering
(Peacock \& Dodds, 1994, Hoyle et al., 1999, Eisenstein \&
Zaldarriaga, 2000).

Below the limits probed by the BCS data, the model predicts a
significant flattening of the luminosity function. This is an
unavoidable consequence of energy injection: in low mass haloes, most
of the gas is ejected, resulting in very low luminosities and a 
`stretching' of the luminosity function in this region.  The space
density of low-luminosity ($L_{\rm X} < 10^{42} h^{-2}\ergs$) systems
is therefore a strong test of this model. The absence of luminous
haloes around spiral galaxies reported by Benson et al.\ (2000)
supports this aspect of the model.

The evolution of the luminosity function is another important test of
the model. The dotted line in Fig~\ref{fig:xlffig} shows the XLF at
$z=0.5$. This evolves very little relative to the present-day
function. The results from the weak evolution of the temperature
function in this cosmological model (Eke et al. 1998) combined with
the weak trend of increasing luminosities with higher redshift at
fixed temperature seen in Fig~\ref{fig:ltfig}.  The model predictions
compare very favourably with the available measurements from deep
ROSAT surveys (eg. Scharf et al. 1997; Rosati et al. 1998; Vikhlinin
et al. 1998; Nichol et al. 1999; Jones et al. 2000) which show no
significant evolution of the luminosity function over the redshift
range $0<z<0.8$. The evolution seen at the bright end is, however,
sensitive to the CDM power spectrum adopted. For instance, the
$\sigma_8=0.73$, $\Gamma=0.07$ model discussed above would imply that
the most massive clusters ($L_{\rm X} > 5\times 10^{44} h^{-2}\ergs$)
should have significantly lower space density at $z=0.5$ than at the
present day. It is unclear whether this is supported by current X-ray
data (see Jones et al.\ 2000 for a discussion).

\subsection{X-ray Emission in the High-Redshift Universe}

\begin{figure}
\begin{centering}
  \psfig{file=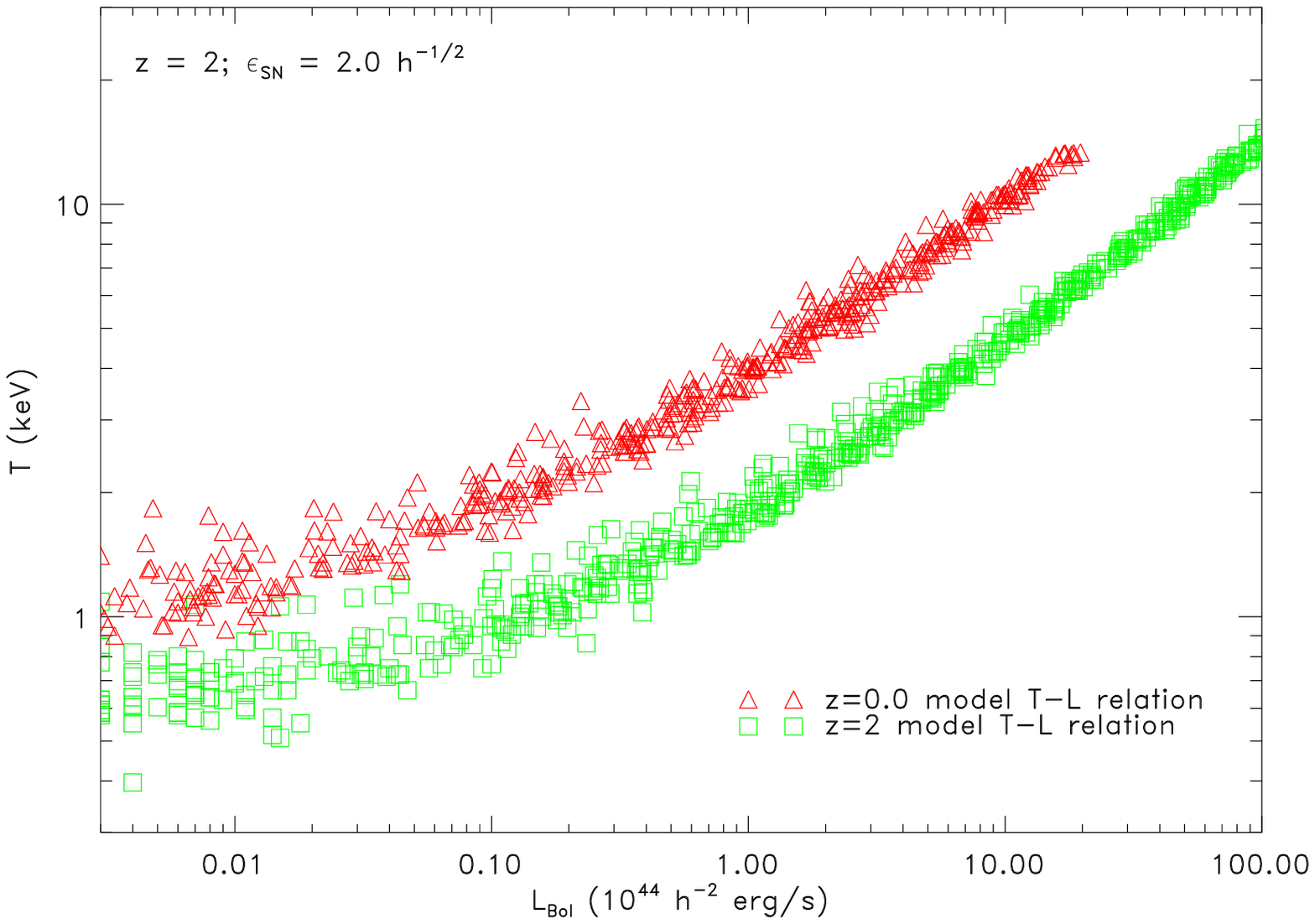,width=8cm}
  \psfig{file=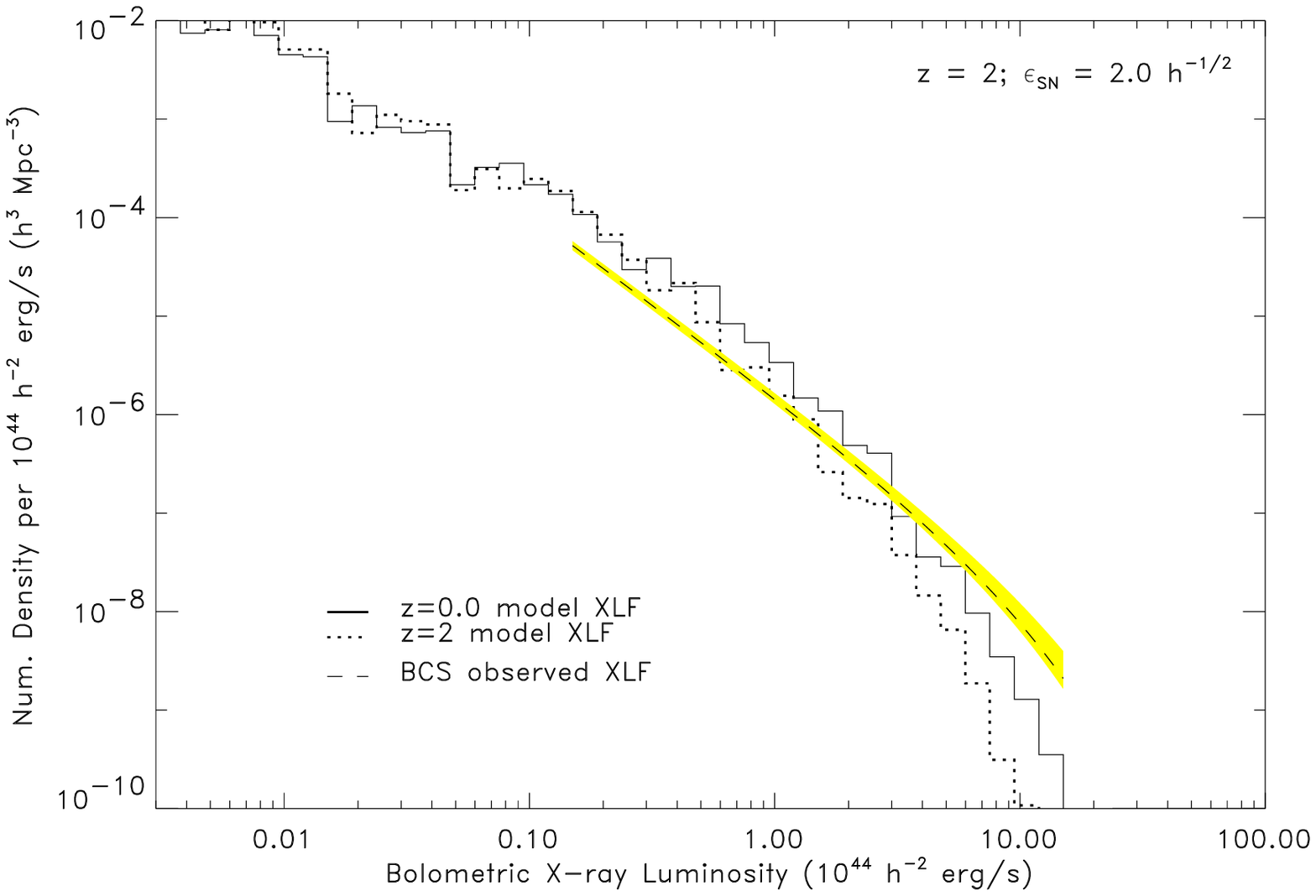,width=8cm}
\end{centering}
\caption{Predictions for the X-ray universe at $z=2$. \emph{Upper
panel:} the T-L relation (triangles: $z=0$, squares:
$z=2$). \emph{Lower panel:} the X-ray luminosity function (solid:
$z=0$, dotted: $z=2$). Both panels assume $\esn=2.0\,h^{-1/2}$. The
observational relations at $z=0$ are also plotted for comparison.}
\label{fig:z2fig}
\end{figure}

We can use the model to predict the evolution of the X-ray emission
from haloes out to high redshifts ($z>2$).  The Cole et al. (2000)
model of galaxy formation and evolution matches reasonably well
observations of the evolution of the cosmic star formation rate over
these long look-back times.  We can thus predict the evolution of the
supernova heating rate out to very high redshift, as is
required in order to model the evolution of the XLF at high redshifts.
We focus on the $\esn=2.0\,h^{-1/2}$ model in what follows, because
this produces the best fit to the present-day XLF.

The model predictions for the T-L relation and the XLF at $z=0$ and $z=2$
are shown in Fig.~\ref{fig:z2fig}.  At a given temperature, high-$z$
clusters are 
substantially more luminous than their present-day
counterparts. However, because of hierarchical clustering, high
temperature systems are increasingly rare at high redshift. At $z=2$,
this decline in abundance offsets the modest increase in X-ray
luminosity at given $T$. As a result, even at $z=2$ the luminosity
function is predicted to be close to that observed at the
present-day. This is consistent with observational limits on the
contribution of clusters to the X-ray background (e.g.  Burg et
al. 1993 and Wu et al. 1999b).

\subsection{The Epoch of Galaxy Formation}

\begin{figure}
\begin{centering}
  \psfig{file=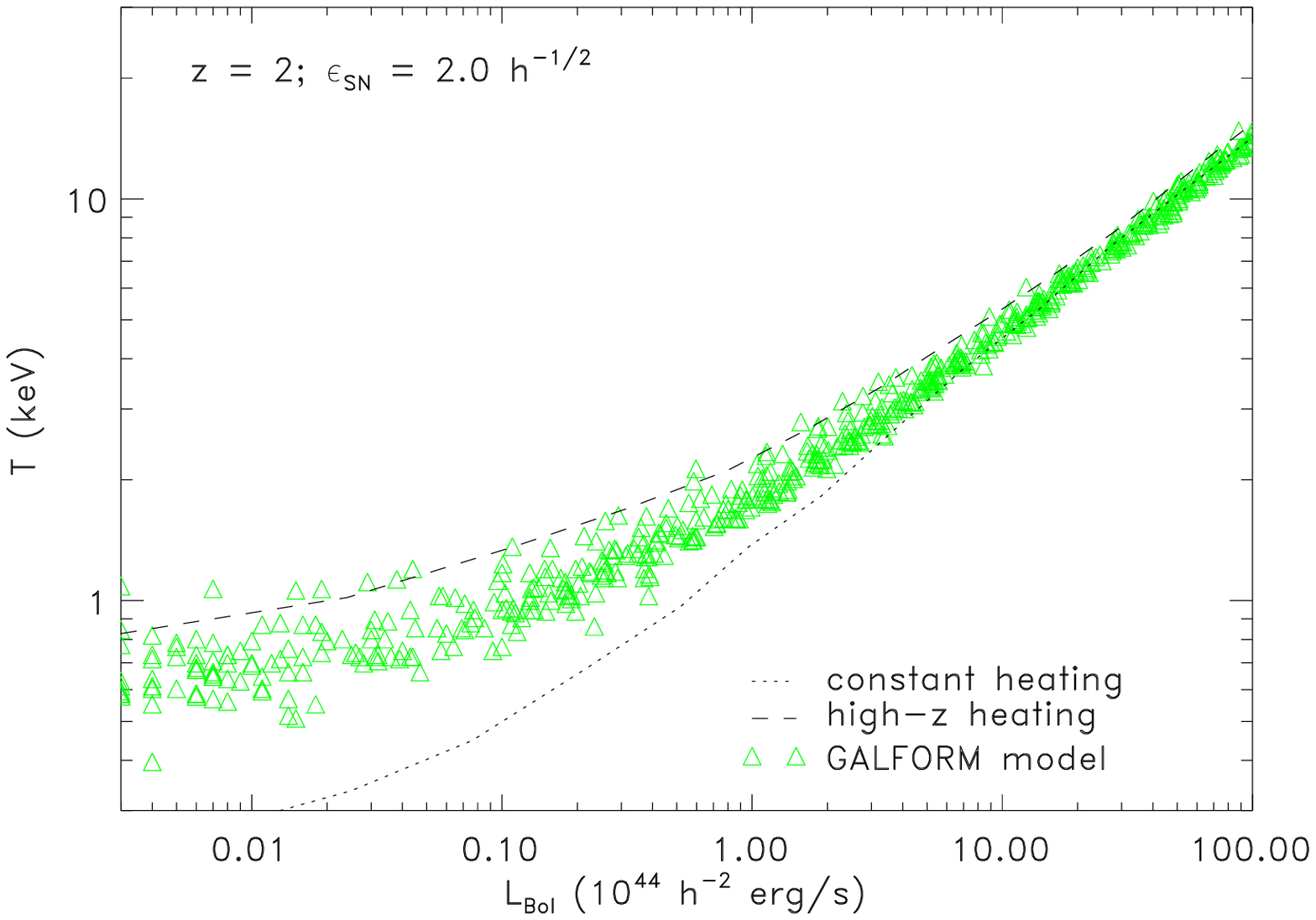,width=8cm}
  \psfig{file=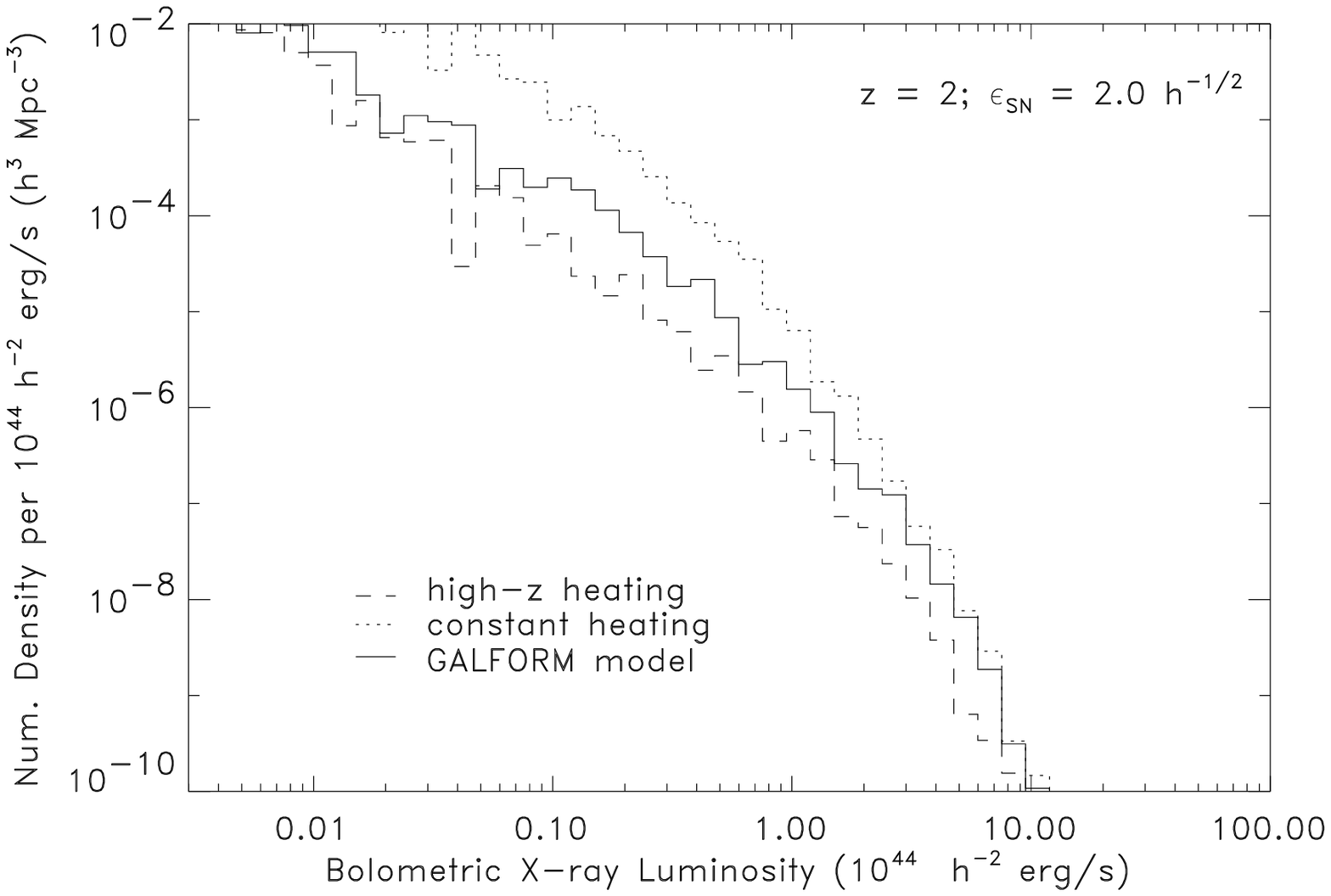,width=8cm}
\end{centering}
\caption{Predicted cluster properties at $z=2$ for different models
for the redshift dependence of the heating. {\it Upper panel}: the
temperature-luminosity relation for the {\sc galform} model compared
to (A)~a model in which the heating occurs at a uniform rate (dotted
line) and (B)~a model in which the heating occurs at high redshift
(dashed line).  {\it Lower panel}: the X-ray luminosity function for
the same models at $z=2$. }
\label{fig:epochfig}
\end{figure}

We have argued that the slope of the temperature-luminosity relation
reflects the energy input from galaxies and AGNs. Now
we examine whether the evolution of clusters can be used to constrain
the epoch at which this heating occurs.  We contrast the {\sc galform}
model (with $\esn=2.0\,h^{-1/2}$) with two simple {\em ad hoc} heating
models. In the first, the heating occurs at a constant rate over
cosmic time (model~A); in the second, the heating occurs only at high
redshift so that the excess energy remains constant below $z=2.0$
(model~B). Model~B is intended to mimic the effect of ``pre-heating''
of the intergalactic medium as in the model proposed by (eg.) Balogh et
al.\ (1999). The total energy injection has been adjusted to match the
present-day XLF of the {\sc galform} model. Models A and B give XLFs
at $z=0$ which are almost identical to that from GALFORM, when
normalized by this procedure.

We contrast these two simple models with our fiducial model based on
hierarchical galaxy formation in Figure~\ref{fig:epochfig}.  The upper
panel shows the median T-L relations predicted by each of the models
at $z=2$. Similar but less pronounced differences exist at $z=1$ and
at $z=0.5$. The models diverge at low luminosities since the relative
effect of the injected energy is greatest for small clusters. It is
not surprising that the differences between the models at the bright
end are small, as the heat input is fairly unimportant for these
clusters.  As expected, the two simple heating models bracket the {\sc
galform} model, although the latter seems closer to model B in
which the heating occurs at high redshift.  This reflects the fact
that galaxy formation in proto-cluster regions is accelerated relative
to that in an average region of the universe, simply due to the higher
density there.

The lower panel in Fig.~\ref{fig:epochfig} shows the differences
between the luminosity functions at $z=2$ for the three models. As
expected from the upper panel, the two simple models again bracket the
behaviour of {\sc galform}.  The luminosity function for the constant
heating model shows stronger positive evolution (ie. a higher number
density at higher redshift) than the model in which the heating has
already occurred before this epoch. These differences offer an
interesting possibility for determining the epoch of galaxy formation:
if it becomes possible to distinguish between different models for the
redshift dependence of the heating using X-ray observations, this will
then provide a strong constraint on the epoch at which most of the
stars in the universe formed.

\section{Discussion}

As we have shown, a model in which the intracluster gas is heated as
galaxy formation proceeds provides a good explanation for the slope of
the T-L relation and for the evolution of the X-ray luminosity
function.  The problem with associating this heating with supernovae
is the large amount of energy that is required, between 1.3 and
$2.0\, \times 10^{49} h^{-1/2}\,\erg$ per solar mass of stars
formed. This corresponds to an energy of 0.6 -- 1.0 keV per particle
in the intracluster medium. This is comparable to the energy injection
requirement (1 -- 2 keV per particle) derived by Wu et al. (1999a),
showing that 
the conclusions about the energetics do not depend greatly on the
details of the heating model. Even with optimistic assumptions about
the supernova rate, this amount of heating would require that the
energy of the supernova explosions  be transferred to the
intergalactic plasma with an efficiency close 
to unity. This seems unrealistic.

An alternative source for heating the ICM is AGNs and quasars. If
AGN activity is closely linked to the fuelling of star formation, then
such activity will effectively enhance the value of $\esn$. In
this case, the effect of AGN heating can easily be incorporated into
our model. Our conclusions would be unchanged apart from the interpretation
of $\esn$. If, on the other hand, the energy input comes
predominantly from the most powerful AGN early in the history
of the universe, it would be more appropriate to treat the energy
injection as a uniform preheating of the gas prior to gravitational
collapse of the dark matter halos. Assuming the energy sources were
sufficiently uniformly distributed, the effects of the heating might
be better modeled by assuming that the gas entropy is raised to some
uniform value before collapse (eg. Evrard \& Henry 1991; Navarro,
Frenk \& White 1995; Bower 1997; Balogh, Babul \& Patton 1999;
Valageas \& Silk 1999). A possible problem of this scheme is the high
temperature it implies for the diffuse IGM. For example, Balogh, Babul
\& Patton (1999) require a temperature of $1.8\times 10^6$~K for a
preheating epoch of $z=3$ in our $\Lambda$CDM cosmology. This is in
stark contrast to the IGM temperature estimated from the Doppler
widths of Ly-$\alpha$ forest lines. For example, Theuns et al (2000)
estimate $T_{\rm IGM} \sim 15,000$ K at this redshift. Thus, unless the
clouds giving rise to the Ly-$\alpha$ forest or the precursor gas of
the ICM are atypical, a model in which the heating occurs within
already virialised haloes seems preferable.

Dark haloes build up by mergers of smaller progenitor halos.  Although
we include energy input from the complete history of star formation in 
each halo, we assume that the way this heating is distributed between the
earlier haloes is unimportant. In particular, we ignore
any dependence of the energy released during the gravitational
collapse of a halo on the distribution of the gas in the progenitor halos. 
If the gas has already
been heated by supernovae or AGN in the progenitor halos, then the
dynamics of the collapse of the gas and its shock heating will be
modified.
It is quite possible that this effect could give rise to an
``amplification'' of an initial energy excess, thus easing the
requirements on the heating efficiency. Another limitation of our
approach is that we have only included the effects of cooling in an
approximate way, by removing from the ICM the gas which should have
cooled to low temperature, based on a calculation which does not
explicitly include the effects of the energy injection. However, the
cooling rates will be modified by the effects of the energy injection,
and conversely some of the excess energy may be lost by radiative
cooling. It is clearly vital to understand all of these processes
better, which can best be done through well-targeted numerical
simulations (eg. Pearce et al. 2000).


Finally, we must recall that galaxy formation and X-ray evolution have
not been treated in a fully self-consistent fashion in this paper. We
have taken the successful {\sc galform} model of galaxy formation,
with the same parameters as in Cole et al. (2000), and used it to
predict the energy injection and thus the evolution of cluster and
group X-ray properties.  In practice, we should use the methods
developed here to calculate the gas density profiles of all haloes at
each epoch, as modified by the energy injection, compute gas cooling
rates using these modified profiles, and then calculate the energy
injection from the resultant star formation histories. This represents
a large computational overhead on the standard {\sc galform} model,
but is clearly an important next step to take.

\section{Conclusions}

In this paper, we have addressed the problem of why the observed
properties of X-ray clusters do not conform to simple scaling
relations. In particular, we have considered why the observed
correlation between X-ray temperature and X-ray luminosity is
significantly shallower than the adiabatic scaling solution, while the
X-ray luminosity function evolves less rapidly than predicted in
popular cold dark matter cosmologies. First, we argued that the
effects of gas cooling in clusters (which break the scaling relations)
do not resolve the problem.  We then considered the heating of the
intracluster gas by the energy released during galaxy formation, by
combining the semi-analytic model of Cole \etal (2000) with a simple
model for the radial profile of the intracluster gas. Our main
conclusions, applicable in the $\Lambda$CDM cosmology, may be
summarized as follows:

\begin{itemize} 
\item Heat input into the intracluster gas by processes associated
with the formation of cluster galaxies, such as supernovae and/or AGN
winds, will flatten the slope of the temperature-luminosity
relation. The combined model gives a reasonable match to the
observations if energy is injected at a level of 1.3--2.0
$\times 10^{49} h^{-1/2}$~ergs per solar mass of stars formed (or,
equivalently, 
0.6--1 keV per particle in the intracluster medium). Values within
this range produce broadly acceptable models, but lower values result
in a better match to groups with $T\approx 1\keV$, while higher values
produce a better match to the most massive clusters.

\item The interplay between the ongoing energy injection from galaxies
and the growth of clusters by hierarchical clustering causes the $T-L$
relation to evolve little to moderate redshifts. This is consistent with
recent data based on ASCA observations.

\item The present-day X-ray luminosity function in the model
approximately matches observations, but the model over-produces low
luminosity groups and under-produces very luminous clusters. Fine
tuning the cosmological parameters or other details of the model may
remove these discrepancies.

\item Similar factors to those that regulate the evolution of the
$T-L$ relation result in only weak evolution of the luminosity
function to $z=0.5$. This too is consistent with current data.

\item The properties of clusters at high redshift provide a test of
the model, since all of the free parameters are fixed to achieve
agreement with present-day data. In particular, the model predicts
little evolution in the X-ray luminosity function even out to
$z=2$. The predicted near constancy of the luminosity function is
testable with the current generation of X-ray satellites.

\item The main difficulty of our model is that it requires an amount
of energy injected per unit mass of stars formed which is comparable
to the total energy available from supernovae. This would require the
supernova explosion energy to couple to the intracluster gas with very
high efficiency, with minimal losses by radiative cooling during the
expansion of the supernova remnants through the ISM of the host
galaxies. However, additional energy sources associated with galaxy
formation may also contribute, such as the mechanical energy liberated
by AGN winds. Alternatively (or additionally), an initial heat input
to the intracluster medium might be amplified during the build-up of
the cluster by mergers. Detailed numerical simulations are required to
test the effectiveness of this process.
\end{itemize}

Our work demonstrates that the shape and evolution of the X-ray luminosity
function and T-L relation are potentially powerful probes of the mode and
efficiency of galaxy formation. Future observations with Newton and Chandra
should be able to test these ideas.

\section*{Acknowledgements}

Thanks to Ed Lloyd-Davies and Trevor Ponman for extensive discussions and to 
Simon White for valuable suggestions. 
This project has made extensive use of Starlink computing facilities
and was supported by the PPARC rolling grant on `Extra-galactic
astronomy and cosmology' at Durham, and by the EC TMR Network on
`Galaxy Formation and Evolution'. CSF acknowledges the support of a
Leverhulme Research Fellowship. CGL acknowledges support at SISSA from
COFIN funds from MURST and funds from ASI.

\appendix

\section{The Effect of Cooling on the T-L Relation}

In the absence of radiative cooling and energy input from galaxy
formation, and assuming that all clusters have density profiles which
are simply rescaled versions of each other, then the bolometric X-ray
luminosities of clusters should vary as
\begin{equation}
L_{\rm X} \propto f_g^2 \rho_{\rm vir} M \Lambda(T)
\end{equation}
where $f_g$ is the fraction of the cluster mass in the form of hot
gas, $\rho_{\rm vir} \propto M/R_{\rm vir}^3$ is the mean total
density within the virial radius, and the cooling rate per unit volume
is proportional to $\rho^2 \Lambda(T)$, with $\Lambda(T) \propto
T^{1/2}$ for bremsstrahlung radiation. Assuming that the density
depends on the collapse redshift $z_{\rm f}$ of the cluster as
$\rho_{\rm vir} \propto \rho_0 (1+z_{\rm f})^3$ ($\rho_0$ being the
present mean density of the universe), and using
equation~(\ref{eq:tvir}) for the temperature, we obtain the scaling
law
\begin{equation}
L_{\rm X} \propto f_g^2 \rho_0^{1/2} (1+z_{\rm f})^{3/2} T^2
\label{eq:lxgrav}
\end{equation}
(eg. Kaiser 1986, 1991; Evrard \& Henry 1991; Bower 1997; Kay \& Bower
1999). We have explicitly included the dependence on the collapse
redshift of the cluster, $z_{\rm f}$, to make it clear that the
scaling depends on this rather than on the redshift at which the
cluster is observed.

As we have discussed, the T-L relation implied by
eqn.~(\ref{eq:lxgrav}) is too steep compared with the observed
luminosities and temperatures of clusters.  Eqn.~(\ref{eq:lxgrav})
suggests that 
the relation might be made shallower if lower temperature clusters had
systematically lower collapse redshifts.  In hierarchical models,
however, smaller mass clusters are expected to collapse at higher
redshifts --- the opposite to the required trend.

In this appendix, we will use simple scaling arguments to argue that
the effects of radiative cooling by itself are not sufficient to bring
the predicted T-L relation into line with the observed one.  The
effects of radiative cooling on the density and temperature profiles,
and thus X-ray luminosities, of spherical clusters have been the
subject of various analytical (e.g. Bertschinger 1989) and numerical
(e.g. Lufkin et al. 2000) investigations. These studies show that outside
the cooling radius $\rcool$, defined such that the local radiative
cooling time is equal to the age of the system, the density and
temperature are almost unchanged from their initial values. Inside
$\rcool$, the gas flows in and then drops out of the flow completely
due to radiative cooling.  In the case that the initial density profile 
is steep, this results in a reduction in the gas
density within $\rcool$, and thus also a reduction in $L_{\rm X}$, On the other hand,
if the initial density profile is very shallow, then the gas density
and $L_{\rm X}$ may be boosted.

Consider first the simple case that the gas density profile is that of
a singular isothermal sphere, $\rho(r) \propto \rho_0 (1+z_{\rm f})^3
(r/R_{\rm vir})^2$, with $T=T_{\rm vir}$. We define $\rcool$ as the
radius where the local cooling time equals the age of the universe
$t_{\rm H}$ (we choose the age of the universe rather than that of the
halo in order to derive the maximum effect of cooling). Thus:
\begin{equation}
 \tcool(\rcool) \propto \rho(\rcool)^{-1} T(\rcool)/\Lambda(T)
        \propto t_{\rm H}.
\end{equation}
(Note that we suppress the dependence on the gas fraction $f_g$ in
this and the following equations.)  The fraction of the gas which is
able to cool is then
\begin{equation}
  \fcool = {\rcool \over R_{\rm vir}} \propto \left( \rho_0
  (1+z_{\rm f})^3 T^{-1/2} t_{\rm H} \right)^{1/2}. 
\label{eq:fcool}
\end{equation}
The self-similar cooling flow solutions of Bertschinger (1989) show
that the density profile flattens to $\rho\propto r^{-3/2}$ within
$\rcool$, so the X-ray luminosity scales as
\begin{equation}
  L_{\rm X} \propto \rcool^3 \rho(\rcool)^2 \Lambda(T) \propto
                { T^2 \rho_0^{1/2} (1+z_{\rm f})^{3/2} \over \fcool}.
\end{equation}
The greater the fraction of gas that is able to cool, the more the
luminosity is reduced below that of
eqn.~(\ref{eq:lxgrav}). Substituting equation~(\ref{eq:fcool}) for
$\fcool$, we then find
\begin{equation}
  L_{\rm X} \propto T^{9/4} t_{\rm H}^{-1/2}.
\label{eq:lxcool}
\end{equation}
This only slightly improves the match to the data compared to the case
of no cooling. For high temperature clusters, the relation becomes
slightly shallower than before, but the effect is not
sufficient. Moreover, in cooler clusters, the emissivity is enhanced
by recombination radiation and the slope of the T-L relation becomes
steeper again.

The above equations apply in the case that the initial gas density
profile is singular. If, as is more realistic, the gas profile
initially has a core of radius $r_c$ within which the density is
constant, then the behaviour is modified. The cooling time will be
constant within the core, so the density profile will remain almost
unchanged until the age of the system is equal to this cooling time,
and $L_{\rm X}$ will approximately follow
equation~(\ref{eq:lxgrav}). As a cooling flow starts up within the
core, the X-ray luminosity may be enhanced, but at most by a factor
of a few. The cooling radius $\rcool$ will then grow beyond the core
radius $r_c$, and $L_{\rm X}$   will converge towards the
behaviour in equation~(\ref{eq:lxcool}). Thus, $L_{\rm X}$ will scale
approximately as (\ref{eq:lxgrav}) for $t_{\rm H}<\tcool(r_c)$ and as
(\ref{eq:lxcool}) for $t_{\rm H}>\tcool(r_c)$, and our previous
conclusions about the T-L relation not being reproduced remain
unchanged. 

An estimate of the effects of cooling based on our $\beta$-model
approach is shown by the dot-dashed line in
Figure~\ref{fig:ltfig}. Starting from the default halo gas profile, we
have calculated the gas mass within the cooling radius for each of the
simulated haloes.  This gas is removed, and the $\beta$-profile is
then adjusted so that the remaining gas mass is distributed within the
virial radius using the same boundary conditions as discussed in \S2,
keeping $r_c$ constant but reducing $\beta$. (If instead $r_c$ is
varied and $\beta$ kept constant, the resulting T-L relation is very
similar.) The T-L relation that results is close to that predicted by
the scaling arguments discussed above, and fails to match the observed
data.

\end{document}